\documentclass[aps,prl,twocolumn,superscriptaddress]{revtex4-1} 
\usepackage{graphicx,subfigure}
\usepackage{amsmath,amssymb}
\usepackage{mathtools}
\usepackage{multirow}
\usepackage{textcomp}
\usepackage{float}
\usepackage{color}
\usepackage{ulem}
\usepackage{dsfont}
\usepackage{pdfpages}

\newcommand{\ket}[1]{\ensuremath{\left\vert{#1}\right\rangle}}

\newcommand{\bracket}[3]{\ensuremath{\langle{#1}\vert{#2}\vert{#3}\rangle}}
\newcommand{\braket}[2]{\ensuremath{\left\langle{#1}\vert{#2}\right\rangle}}

\newcommand{\abs}[1]{\ensuremath{\left\vert{#1}\right\vert}}

\newcommand{\G}{\ensuremath{\mathcal{G}}}

\renewcommand{\vec}[1]{\ensuremath{\mathbf{#1}}}

\newcommand{\Heff}{\ensuremath{H_\mathrm{eff}}}

\newcommand{\tdet}{\ensuremath{t_d}}

\newcommand{\Gom}{\ensuremath{g_0}}
\newcommand{\omegam}{\ensuremath{\Omega_\mathrm{M}}}
\newcommand{\np}{\ensuremath{n}}

\newcommand{\Hin}{\ensuremath{H_\mathrm{in}}}

\newcommand{\tin}{\ensuremath{t}}

\makeatletter
\AtBeginDocument{\let\LS@rot\@undefined}
\makeatother

\begin{document}
\title{Painting Non-Classical States of Spin or Motion with Shaped Single Photons}
\author{Emily J. Davis}
\affiliation{Department of Physics, Stanford University, Stanford, California 94305, USA}
\author{Zhaoyou Wang}
\affiliation{Department of Applied Physics, Stanford University, Stanford, California 94305, USA}
\author{Amir H. Safavi-Naeini}
\affiliation{Department of Applied Physics, Stanford University, Stanford, California 94305, USA}
\author{Monika H. Schleier-Smith}
\affiliation{Department of Physics, Stanford University, Stanford, California 94305, USA}

\begin{abstract}
We propose a robust scheme for generating macroscopic superposition states of spin or motion with the aid of a single photon.  Shaping the wave packet of the photon enables high-fidelity preparation of non-classical states of matter even in the presence of photon loss.  Success is heralded by photodetection, enabling the scheme to be implemented with a weak coherent field.  We analyze applications to preparing Schr\"{o}dinger cat states of a collective atomic spin or of a mechanical oscillator coupled to an optical resonator.  The method generalizes to preparing arbitrary superpositions of coherent states, enabling full quantum control.  We illustrate this versatility by showing how to prepare Dicke or Fock states, as well as superpositions in the Dicke or Fock basis.
\end{abstract}
\date{\today}

\maketitle
Macroscopic quantum superposition states theoretically have wide-ranging applications in precision measurement \cite{bollinger1996optimal,kessler2014heisenberg}, quantum error correction \cite{leghtas2013hardware}, continuous-variable quantum communication \cite{van2001entangled}, and tests of fundamental physics \cite{romero2011large,asadian2014probing,arndt2014testing,pikovski2015universal}.  To date, Schr\"{o}dinger cat states have been prepared in groundbreaking experiments with optical \cite{ourjoumtsev2006generating, sychev2017enlargement} and microwave \cite{deleglise2008reconstruction,vlastakis2013deterministically,wang2016schrodinger} photons and in chains of trapped ions \cite{leibfried2005creation,monz201114}.  Increasingly macroscopic cat states of matter could advance the stability of atomic clocks to fundamental quantum limits \cite{kessler2014heisenberg,eldredge2016optimal} or elucidate the interplay of quantum mechanics and gravity \cite{romero2011large,arndt2014testing,asadian2014probing,pikovski2015universal}.

One approach to generating Schr\"{o}dinger's cat states that has been proposed in diverse contexts employs an ancilla qubit, mapping a superposition of the qubit into a superposition of two coherent states of a collective spin \cite{muller2009mesoscopic,saffman2009efficient} or microwave field~\cite{brune1992manipulation,vlastakis2013deterministically,wang2016schrodinger,deleglise2008reconstruction}. A challenge is that the ancilla---e.g., a photon \cite{jiang2008anyonic} or Rydberg atom \cite{muller2009mesoscopic,saffman2009efficient,brune1992manipulation}---is generally subject to dissipation.  In dissipative systems, heralded schemes \cite{duan2001long,lund2004conditional,mcconnell2013generating,galland2014heralded,christensen2014quantum,mcconnell2015entanglement,casabone2013heralded,riedinger2016non,welte2017cavity} can generate highly pure non-classical states that require much stronger coupling to access deterministically \cite{barontini2015deterministic,o2010quantum}.  A particularly versatile scheme proposed by Chen \textit{et al.} ``carves'' a many-atom entangled state from a simple initial state via a quantum non-demolition measurement with a single photon \cite{chen2015carving}.  Still, the measurement fidelity is fundamentally limited by finite interaction strength relative to the photon loss rate.


Here, we propose a heralded scheme for generating macroscopic superposition states with high fidelity even at finite interaction-to-decay ratio, using only modulated laser light and a single-photon detector.  Our scheme employs a photon as a ``brush'' for painting superpositions of coherent states at designated points in the phase space of a collective atomic spin or a mechanical oscillator.  The phase space points are selected by shaping the time dependence of the photon pulse.  The approach generalizes to
selecting multiple points or a continuous curve in phase space, enabling full quantum control.  

\begin{figure}[htb]
\includegraphics[width=\columnwidth]{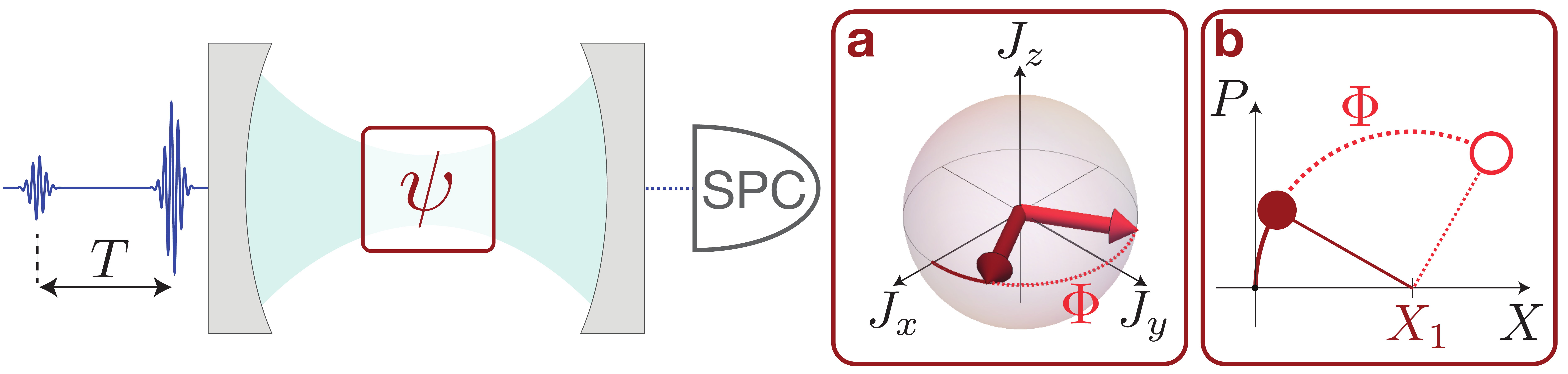}
\caption{\textbf{Painting Schr\"{o}dinger's cat states with shaped single photons}. A cat state of \textbf{(a)} spin or \textbf{(b)} motion, with phase separation $\Phi$, is generated by a photon in a superposition of two pulses separated by a time $T = \Phi/\Omega_{S/M}$. The single photon is introduced by driving a cavity with a weak coherent field.  Success is heralded by detection of the photon by the single-photon counter (SPC) at the cavity output. \label{fig:overview}}
\end{figure}


Our approach is illustrated in Fig. \ref{fig:overview}, where we first show how to prepare the collective spin $\vec{J}$ of an atomic ensemble in a superposition of two distinct orientations (Fig. \ref{fig:overview}a).  To manipulate the ensemble, we employ a dispersive atom-light interaction
\begin{equation}\label{eq:Hs}
H_S = \Omega_S c^\dagger c J_z,
\end{equation}
where $c^\dagger c$ represents the number of photons in an optical resonator mode, $J_z$ denotes the population difference between two internal states, and $\Omega_S$ denotes the differential ac Stark shift due to a single photon; we set $\hbar=1$.  After initializing the ensemble in a coherent spin state $\ket{\psi_0}$ along $\vec{x}$, we let the atoms interact with a photon in a wavepacket consisting of two short pulses at times $t = 0$ and $t = T$.  If we detect a photon exiting the cavity at a time $\tdet >T$, it may have interacted with the atoms for either a time $\tdet$ or a time $\tdet-T$.  The ensemble is thus projected into a superposition
\begin{equation}
\label{eq:kitten_S}
\ket{\psi_1(t_d)} = c_0\ket{\Omega_S \tdet} + c_T \ket{\Omega_S \tdet-\Phi}
\end{equation}
of two rotated copies $\ket{\varphi} \equiv e^{-i\varphi J_z}\ket{\psi_0}$ of the initial state with angular separation $\Phi = \Omega_S T$.

The amplitudes $c_0$ and $c_T$ of the superposition state depend on the strengths of the two pulses and on their separation time $T$.  For equal pulse strengths, a photon detected at a late time $\tdet > T$ is more likely to have arrived in the second pulse than in the first by a factor $e^{\kappa T}$, where $1/\kappa$ is the cavity lifetime.  However, compensating with unequal pulse strengths will allow for preparing an equal superposition state even in the presence of loss.

The same method can generate a cat state of the motion of a mechanical oscillator initialized in its ground state $\ket{\psi_0}$.  We consider a Hamiltonian
\begin{equation}\label{eq:Hm}
H_M =  \frac{1}{2}\omegam (P^2 + X^2) - \Gom c^\dagger c X,
\end{equation}
where $X = x/x_0$ denotes the displacement normalized to the zero-point length $x_0 = \sqrt{\hbar/(m\omegam)}$ for mass $m$ and frequency $\Omega_M$; $P$ is the conjugate momentum; and $\Gom$ is the optomechanical coupling strength.  If a photon enters the cavity, it displaces the equilibrium position of the oscillator by an amount $X_1 = |\Gom/\omegam|$, and the coherent state $\ket{\psi_0}$ begins to rotate about the new equilibrium position in phase space (Fig. \ref{fig:overview}b).  Thus, a photon entering at a superposition of times $t=0$ and $t=T$, and detected at time $\tdet > T$, projects the oscillator into the superposition of coherent states shown in Fig. \ref{fig:overview}b.  The angular separation $\Phi = \omegam T$ on a circle of radius $X_1$ results in a phase-space separation $2X_1\sin(\Phi/2)$, which can be large for $\Gom > \omegam$.

Imparting a well-defined phase shift $\Phi$ with a single photon requires that the photon pulse be short in time or, equivalently, broad in frequency compared to the uncertainty in the $J_z$- or $X$-dependent cavity frequency.  To nevertheless obtain an appreciable heralding probability, we should drive the cavity with a coherent pulse strong enough to produce a small probability for a single photon to enter the cavity, but weak enough to avoid two-photon events.  To ensure that we detect only photons that have interacted with the system 
, and not the reflected component of the input field, we consider a two-sided cavity driven from one end, with a detector at the far end.


In the limit of a weakly transmissive input mirror, the driving of the cavity is described by a Hamiltonian
\begin{equation}
\Hin(t) = \mathcal{E}(t) c^\dagger + \mathcal{E}^*(t)c.
\end{equation}
The conditional evolution is governed by the non-Hermitian Hamiltonian \cite{gardiner2004quantum}
\begin{equation}\label{eq:Hnh}
\Heff = \Hin + H_{S/M} + \left(\omega_c - i \frac{\kappa}{2}\right)c^\dagger c,
\end{equation}
where  $H_{S/M}$ describes the spin/mechanical system and its interaction with the intracavity light (Eq. \ref{eq:Hs} or Eq. \ref{eq:Hm}); $\omega_c$ is the frequency of the bare cavity mode; and $\kappa$ is the cavity linewidth. Conditioned on transmission of exactly one photon, at time $t_d$, the final heralded state of the system is
\begin{equation}\label{eq:psiE}
\ket{\psi_1} = \bracket{0}{\sqrt{\kappa}\hat{c}\,\mathcal{\hat{T}}e^{-i\int_0^{\tdet}\Heff(t)\,dt}}{0}\otimes\ket{\psi_0},
\end{equation}
where $\ket{0}$ denotes the vacuum state of the cavity and $\mathcal{\hat{T}}$ is the time-ordering operator \cite{SM}.

To analyze the conditional evolution, we let $H_\np = \langle \np\vert H_{S/M} \vert \np \rangle$ denote the Hamiltonian projected onto the subspace with $n$ photons in the cavity.  For the spin system, $H_\np$ generates a precession
\begin{subequations}
\begin{equation}
U_n(t) \equiv e^{-i H_n t} = e^{-i \np \Omega_S J_z t} 
\end{equation}
by an angle proportional to the intracavity photon number.  For a mechanical oscillator, $H_n$ generates a phase-space rotation
\begin{equation}
U_n(t) \equiv e^{-i H_n t} = e^{-i \omegam (a^\dagger-X_n)(a-X_n)t} e^{i \omegam X_n^2 t},
\end{equation}
\end{subequations}
about a point $X_n = n\Gom/\omegam$ that depends on photon number, where $a$ is the annihilation operator for phonons in the mechanical resonator.  We assume an input field sufficiently weak that at most one photon enters the cavity ($n=0$ or $1$).  In this limit, the dynamics of the mechanics are analogous to those of the spins: if the oscillator is initialized in the vacuum state, then in either system, $U_\np$ generates a non-trivial rotation  $U_1(\varphi/\Omega)$ if and only if there is one photon in the cavity.



Drawing on the principle of Fig. \ref{fig:overview}, we will apply these light-induced rotations to prepare target superposition states of the generic form
\begin{equation}\label{eq:psif}
\ket{\psi_*} = \int_0^{\varphi_\mathrm{max}}d\varphi f(\varphi) U_1(\varphi/\Omega)\ket{\psi_0},\\
\end{equation}
with $\varphi_\mathrm{max} \leq 2\pi$. The coefficients $f(\varphi)$ will be determined by the shape of the input pulse $\mathcal{E}(t)$.
If we apply a weak input field $\mathcal{E}(\tin)=\mathcal{E}_0(\tin)e^{-i\omega_c \tin}$ for times $t\geq 0$, detecting a photon at time $t_d$ projects the system into a state
\begin{equation}
\label{eq:psi_t1}
\ket{\psi_1} = \sqrt{\kappa}\int_0^{\tdet} d\tau\,\mathcal{E}_0(\tdet-\tau)e^{-\kappa \tau/2}U_1(\tau)\ket{\psi_0}.
\end{equation}
Here, the integral is over the photon's duration $\tau$ in the cavity.  The exponential decay reflects the fact that the photon is unlikely to have entered long before detection.


Comparing the heralded state $\ket{\psi_1}$ with the target state $\ket{\psi_*}$, we choose a pulse shape
\begin{equation}
\label{eq:drive_f}
\mathcal{E}_0(t)= \epsilon f(\varphi_\mathrm{max}-\Omega t) e^{- \kappa t/2}.
\end{equation}
Here, $\epsilon$ parameterizes the field strength and must satisfy $\epsilon \ll \Omega$ to ensure that at most one photon interacts with the system.  The exponential decay compensates for the finite cavity lifetime and hides all information about when the photon entered the cavity.  Thus, the pulse shape in Eq. \ref{eq:drive_f} produces a heralded state 
\begin{equation}
\ket{\psi_1} = \frac{\sqrt{\kappa}\epsilon}{\Omega} e^{-\kappa t_d/2} U_1(t_d - \varphi_\mathrm{max}/\Omega)\ket{\psi_*},
\end{equation}
equivalent to the target state $\ket{\psi_*}$ up to an overall rotation. 

To produce a \textit{Schr\"{o}dinger cat state}, our derivation confirms that the input field should consist of a pair of short pulses.  Following Eq. \ref{eq:drive_f}, we set

\begin{equation} \label{eq:cat_state}
\mathcal{E}(t) = \frac{\epsilon}{\sqrt{2}\Omega} \left[\delta(t) + \delta(t-T)e^{i\phi}\right]e^{-i\omega_c t - \kappa t/2}.
\end{equation}
In practice, $\delta(t)$ represents a pulse so short that the transmission amplitude for a single pulse would be independent of the system state ($J_z$ or $X$), corresponding to a bandwidth $B \gg \Omega \sqrt{N}$ for an $N$-atom spin ensemble or $B \gg g_0$ for the mechanical oscillator (Fig. \ref{fig:overview}c). The two pulses interfere to produce a transmission amplitude that does depend on the spin or motional state, entangling the output field with the system \cite{chen2015carving}.  Conditioned on detecting a photon at a time $t_d>T$, the system is projected into an equally weighted superposition of two coherent states, as in Eq. \ref{eq:kitten_S} with $c_T = e^{i\phi}c_0$.

By extension, shaping the time dependence of the input field allows for ``painting'' more general superpositions of coherent states, with amplitudes specified by $f(\varphi)$.  Note that the description $f(\varphi)$ for a given target state is not unique, because the coherent states form an over-complete basis.  As a further consequence, although the phase-space trajectory specified by $f(\varphi)$ is restricted to lie on a circle, arbitrary target states $\ket{\psi_*}$ are accessible by painting along such a path.

To provide a recipe for full quantum control, we expand the state of the system in the basis of Dicke states $\ket{J_z = m}$  or displaced Fock states $\ket{\tilde{a}^\dagger \tilde{a} = m}$.  In the latter case, we let $m$ denote the phonon number when the equilibrium position is displaced by a single intracavity photon by defining $\tilde{a} = a - X_1$.  In the expansions
\begin{align}\label{eq:DickeExpansion}
\ket{\psi_0} \equiv \sum_m c^0_m \ket{m},\,\,\,\,\,\,\,
\ket{\psi_*} \equiv \sum_m c_m^f \ket{m}
\end{align}
of the initial and final states, the coefficients are related by $c^f_m = c^0_m f_m$, where $f_m = \int_0^{\phi_\mathrm{max}}d\varphi f(\varphi) e^{i m\varphi}$ is the Fourier transform of the weighting function $f(\varphi)$ in Eq. \ref{eq:psif}.  Thus, to prepare $\ket{\psi_*}$, we apply an input field
\begin{equation}\label{eq:Eforcf}
\mathcal{E}(t) = \frac{\epsilon}{2\pi} \sum_m \frac{c^f_m}{c^0_m}e^{-i [\omega_c + \Omega(m-\mu)] t - \kappa t/2} \,\,\, \mathrm{for} \,\,\, t\in[0,T],
\end{equation}
where $ \mu=0 $ for the spin case and $ \mu = X_1^2 $ for the mechanical oscillator.  The field is applied up to time $T = \varphi_{max}/\Omega \leq 2\pi/\Omega$, with $\mathcal{E}(t) = 0$ for $t \notin [0,T]$.  The target state is theoretically accessible from any initial state in which $c_m^0\neq 0$ whenever $c_m^f \neq 0$.   

A \textit{Dicke state} $\ket{\psi_*}=\ket{m}$ offers an illuminating example.  The state $\ket{m}$ is prepared by a photon of center frequency $\omega_c + \Omega m$ in a pulse of duration $2\pi/\Omega$ with decaying intensity.  The frequency is chosen so that the field is transmitted through the cavity only if $J_z = m$ \cite{chen2015carving}.  The decaying intensity is best understood by considering the measurement back-action, namely, the spin rotation due to the ac Stark shift $\Omega$.  Since the Dicke state is symmetric under rotations about $J_z$, the spin should have equal probability $\propto \abs{f(\varphi)}^2$ of being rotated by any angle $0<\varphi\leq 2\pi$, conditioned on detection of the photon.  To ensure that the detection time $\tdet > 2\pi/\Omega$ provides no information about when the photon entered the cavity, the intensity of the drive must decay at rate $\kappa$.

A \textit{Fock state of motion} can be prepared similarly. When a photon enters the cavity, it exerts a force that drives the system along an arc of radius $X_1$ in phase space. An exponentially shaped pulse of length $2\pi/\Omega$ ensures that this arc is equally likely to end at any point on a circle, thus painting the circular quasiprobability distribution of a Fock state $\ket{m}$. Choosing a center frequency $\omega_c + \Omega (m-X_1^2)$ ensures that the photon can enter the cavity and remain there until time $T$ only by exciting the $m^\mathrm{th}$ motional sideband.

The scheme for preparing Fock states bears a superficial resemblance to a method demonstrated in recent experiments \cite{riedinger2016non,riedinger2018remote}.  There, a single-phonons Fock states are generated by driving the optical cavity at a frequency $\omega_c+\Omega$ and detecting only photons emitted at frequency $\omega_c$. By contrast, the paintbrush method obviates filtering of the optical field emanating from the cavity---a technically limiting aspect in experiments to date.

Moreover, a coherent superposition $\ket{\psi_+} = \frac{1}{\sqrt{2}}(c_0^f\ket{0}+ c_1^f\ket{1})$ can be generated with the same technique, which  allows for encoding arbitary qubit states in the oscillator.  For example, to prepare the equal superposition $c_0^f = c_1^f = \frac{1}{\sqrt{2}}$ starting from the undisplaced vacuum, we require a drive field
	\begin{align}
\mathcal{E}(t) = \epsilon A e^{-i(\omega_c - X_1^2 \omegam)t - \kappa t / 2} \left(X_1 + e^{-i \omegam t} \right).
	\end{align}
where $A = e^{X_1^2/2}/(2\pi\sqrt{2}X_1)$. The engineered driving is not significantly more complex than what is needed to generate a Fock state, demonstrating the versatility of the paintbrush technique.


In principle, even in a lossy cavity, shaping the input pulse according to the loss rate $\kappa$ enables heralded preparation of arbitrary target states with perfect fidelity.  In practice, the heralding rate must compete with the dark count rate of the photodetector.  At the same time, the input field must be weak enough to ensure that the detected photon is the only one that has interacted with the system.  To analyze this trade-off, we consider the general case where the input field is not necessarily weak.

The heralded state for arbitrary drive strength (Eq. \ref{eq:psiE}) can be evaluated analytically for the spin system \cite{SM} or numerically for the oscillator.  We define the \textit{success rate} $R_s(t) = \langle\psi_1\vert\psi_1\rangle$ as the probability per unit time that a single photon is transmitted after time $T$ and \textit{no other photons} are transmitted in the same trial.  In this case, the target state is prepared with fidelity $F_\epsilon = \abs{\langle \psi_1 \vert \psi_* \rangle}^2/\braket{\psi_1}{\psi_1}$ in the absence of dark counts. \
With increasing drive strength, $F_\epsilon$ decreases more slowly than does the success rate $R_s$, so in practice the fidelity is limited by effects of imperfect detection. 

Two practical limitations are finite quantum efficiency $Q$ and the dark count rate $R_d$ of the detector.  Accounting for these effects, a lower bound on the fidelity of the state heralded by a detector click at time $t>T$ is \cite{SM}

\begin{equation}
F_\mathrm{min}(t) = \frac{F_\epsilon(t) R_s(t)}{R_t(t) + R_d/Q},
\label{eq:fidelity}
\end{equation}
where $R_t(t) = \kappa\langle c^\dagger(t) c(t)\rangle$ is the transmission rate.  In the weak drive limit, $R_t = R_s$, but $R_s$ decreases for increasing drive strength because of multi-photon events, reducing the fidelity.  For spins driven by the field in Eq. \ref{eq:Eforcf}, $R_t(t) =  \kappa\abs{\epsilon/\Omega}^2 e^{-\kappa t}$, while $R_s/R_t \approx e^{-\abs{\epsilon/\Omega}^2}$ \cite{SM}.

\begin{figure}[tb]
\includegraphics[width=\columnwidth]{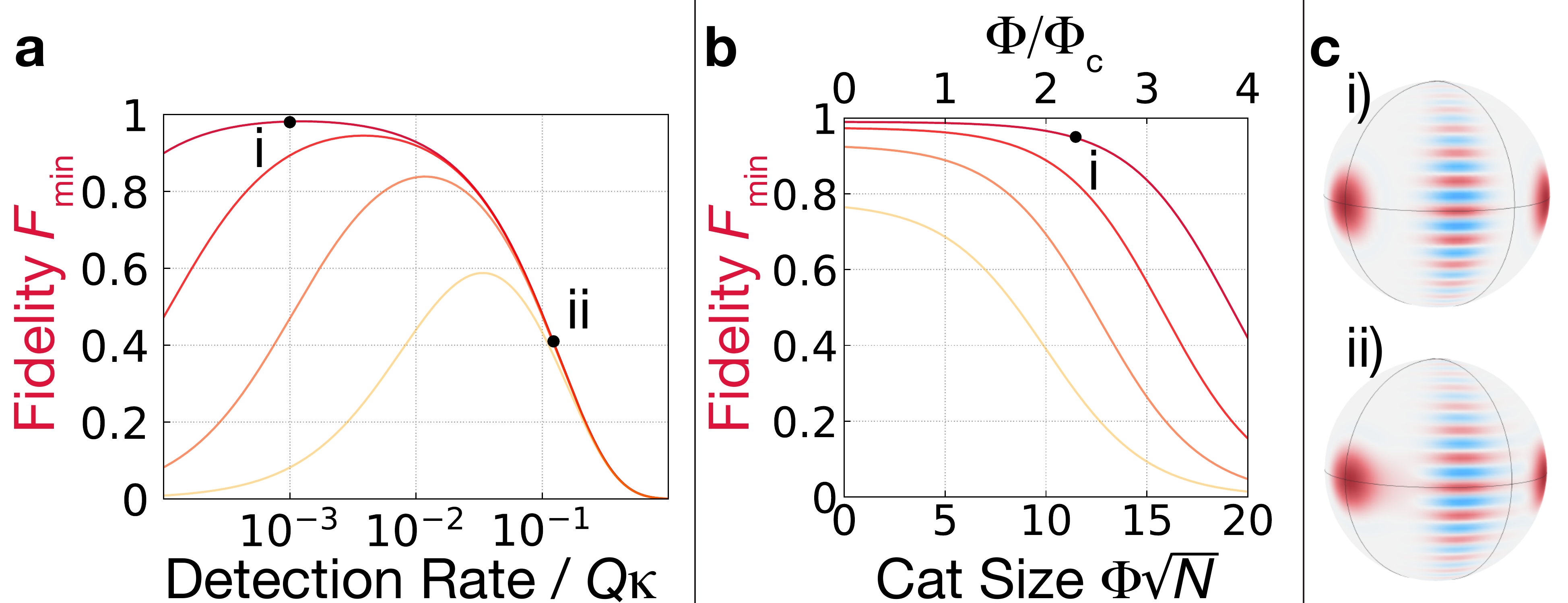}
\caption{\textbf{Fidelity of  spin cat states} at dark count rates $R_d/(Q\kappa) = 10^{-5}, 10^{-4}, 10^{-3}, 10^{-2}$ (dark red to light yellow). {\bf (a)} $F_{\text{min}}(T)$ vs. detection rate $QR_t(T)$ for $\Omega_S = \kappa$, $T=(2\pi/3)\kappa$, and $\Phi = 2\pi/3$. {\bf (b)} $F_\mathrm{min}(T)$ vs cat size $\Phi\sqrt{N}$ or $\Phi/\Phi_c$ at $\eta=50$ for optimum drive strength. Black dots correspond to states illustrated in \textbf{c}(i-ii) at $N=30$, $\Phi = 2\pi/3$.  At high drive strength $\epsilon = \Omega_S$ (ii), undetected transmitted photons cause mild dephasing. \label{fig:success_fidelity}}
\end{figure}

Figure \ref{fig:success_fidelity}(a) shows the dependence of fidelity on drive strength and dark count rate for spin cat states of angular separation $\Phi = \Omega_S T = 2\pi/3$.  At low dark count rate $R_d$, the fidelity is near unity in the weak-drive limit, at the expense of a low detection rate $Q R_t\propto \abs{\epsilon}^2$.  With increasing dark counts, the optimum drive strength increases, and multi-photon events begin to reduce the fidelity. High fidelity is attainable for $R_d \ll Q\kappa e^{-\kappa T}$, a condition easily satisfied if the time $T$ required to rotate the spins is not much longer than the cavity lifetime.


Finite atom-light coupling strength limits the rotation induced by a single photon \cite{SM}.  Specifically, the dispersive coupling is accompanied by absorption that broadens the cavity linewidth to $\kappa_N$ for $N$ atoms, reducing the single-photon phase shift imparted within the cavity lifetime to $\Phi_c = \Omega_S/\kappa_N$.  A fundamental limit $\Phi_c \leq \sqrt{\eta/(2N)}$ is set by the single-atom cooperativity $\eta = \G^2/(\kappa \Gamma)$, where $\G$ is the vacuum Rabi frequency and $\Gamma$ is the linewidth of the atomic transition to which the cavity couples.  Rotations larger than $\Phi_c$ occur only at an exponentially decaying success rate.

The effect of finite cooperativity is illustrated in Fig. \ref{fig:success_fidelity}(b).  The maximum cat size attainable with high fidelity, in units of the coherent state width, is roughly $\Phi_c\sqrt{N}=\sqrt{\eta/2}$.  Yet the attainable cat size furthermore depends logarithmically on $Q\kappa/R_d$ and thus is enhanced by the fact that $Q\kappa\sim 10^3-10^6$/s can be orders of magnitude higher than the dark count rate.  For $Q\kappa/R_d=10^5$, a spin cat of size $\Phi\sqrt{N}=11$ can be prepared with 95\% fidelity at cooperativity $\eta=50$.

\begin{figure}[htb]
\includegraphics[width=\columnwidth]{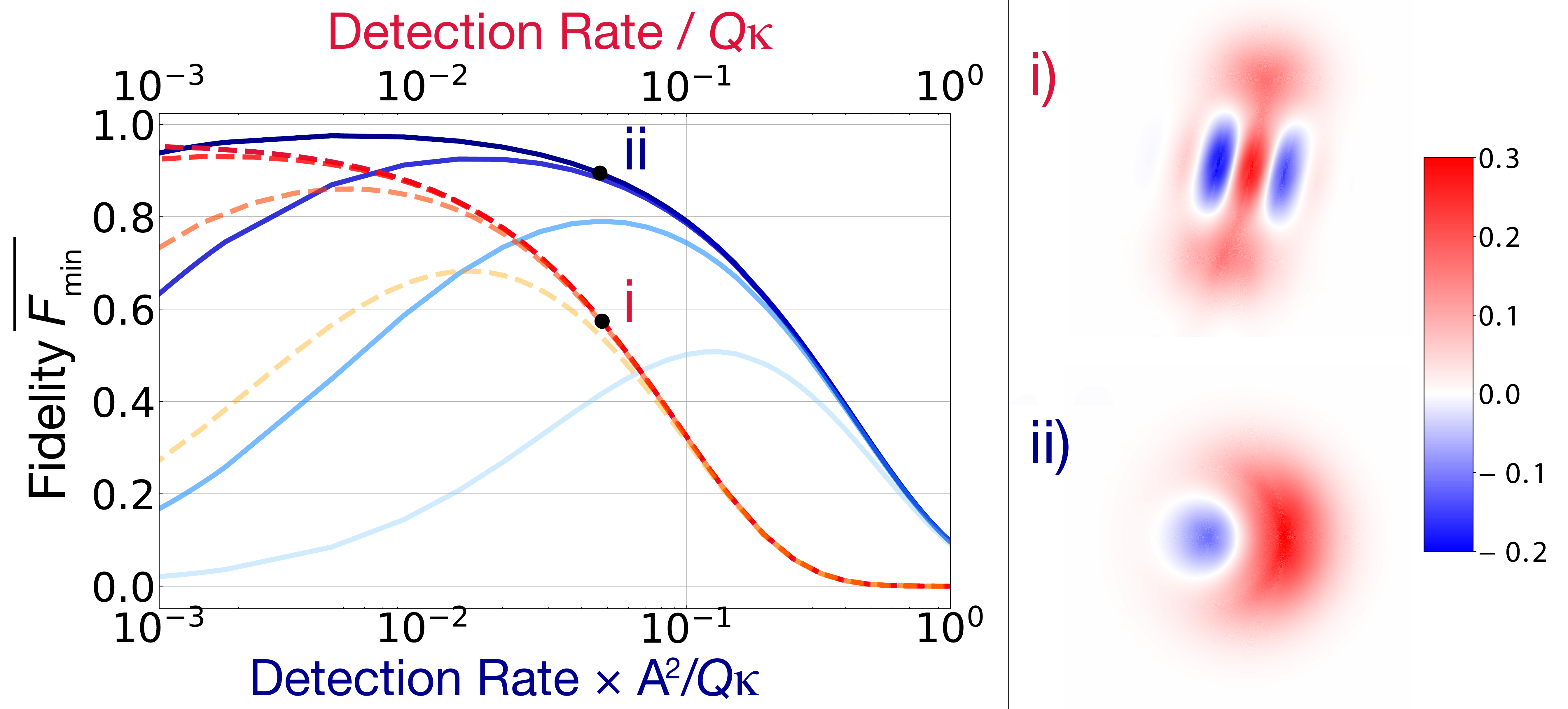}
\caption{\textbf{Non-classical states of motion.} {\bf (i)}  Schr\"{o}dinger cat state of atomic motion for $g_0=\kappa=2\pi\times 100\text{kHz},\Omega_M=g_0/8$ and $T=3/\kappa$.  Dashed red curves show $F_{\text{min}}$ averaged over detection times $T < t_d  < T + 2/\kappa$ for dark count rates $R_d/(Q\kappa) = 10^{-6},10^{-5}, 10^{-4}, 10^{-3}$ (dark to light). {\bf (ii)} Qubit state $\ket{\psi_+}$ of a mechanical oscillator. Solid blue curves show $F_\mathrm{min}$ for $R_d/(Q\kappa) =10^{-10},10^{-9}, 10^{-8}, 10^{-7}$ (dark to light) for $\kappa = 2\pi\times 500~\text{MHz},\Omega_M=2\pi\times 4~\text{GHz},g_0=2\pi\times 1~\text{MHz}$; the detection rate is suppressed by the factor $A^{-2}\approx 8\pi^2 X_1^2$. \label{fig:success_fidelity_OM}}
\end{figure}

For motional cat states generated using the double-pulse sequence in Eq.~\ref{eq:cat_state}, the cat size is at most $\sim\Gom / \omegam$. To produce this separation, the two pulses must have an amplitude ratio on the order of $e^{-\pi\kappa/2\omegam}$, leading to an exponential suppression of count rates as $\kappa$ becomes larger than $\omegam$. Making $\kappa$ and $\omegam$ approximately equal, we find that $\Gom > \kappa$ is required to generate large cat states.  Small motional cat states could be prepared in current atom optomechanics experiments harnessing a Bose-Einstein condensate as a low-mass oscillator to achieve $\Gom\approx\kappa$ \cite{murch2008observation}.  Figure \ref{fig:success_fidelity_OM} shows figures of merit (red curves) for generating a state separated by three times the coherent-state width (Wigner function in Fig. \ref{fig:success_fidelity_OM}.i) with $\kappa_N = \Gom = 8\omegam$ \cite{SM}.

Painting arbitrary superpositions of Fock states (Eq.~\ref{eq:Eforcf}) poses requirements on the system rates similar to those for preparing cat states. The success rate is restricted by the overlap between the displaced Fock state $ \ket{\tilde{a}^\dagger \tilde{a} = m} $ and the undisplaced vacuum, thus scaling as $X_1^{2m} $ for small $X_1 = \Gom / \omegam $.  In this regime, it quickly becomes impractical to access large Fock states. Similarly, the rapid suppression of count rates for large $\kappa/\omegam$ ratios makes it preferable to have $\kappa$ and $\omegam$ on the same order. Complex states with many phonons can be generated efficiently when $\Gom>\kappa$.


In near-term experiments, a weaker optomechanical coupling $\Gom < \kappa$ suffices to paint the mechanical qubit state $\ket{\psi_+}$, illustrated in Fig. \ref{fig:success_fidelity_OM} for parameters similar to those in Ref.~\cite{riedinger2018remote}.  Despite the suppression of the success rate by $X_1^2 = (g_0/\omegam)^2\sim 10^{-7}$, the fast cavity bandwidth enables a heralding rate $\sim 800$/s.  Preparing similar states in other demonstrated optomechanical systems, such as membrane-in-the-middle~\cite{Purdy2015,Underwood2015} ($X_1^2\sim 10^{-9}$) or superfluid resonators~\cite{Shkarin2017}  ($X_1^2\sim 10^{-10}$), may also be possible if dark counts can be sufficiently suppressed.



We have demonstrated how to prepare arbitrary target states of spin or motion using a robust single-photon heralding scheme.  A weak, time-shaped coherent pulse of light enables high-fidelity preparation of non-classical states even in the presence of photon loss.  The generation of cat states can be made more deterministic by driving an ancilla qubit to emit a single time-shaped photon into the cavity.  Extended to multiple spatially separated cavities, the painting scheme could generate long-distance entangled states in quantum networks \cite{komar2014}.




\begin{acknowledgments}
E.D. and Z.W. contributed equally to this work. E.D. acknowledges support from the NSF and Hertz Foundation. M.S.-S. acknowledges support from the AFOSR and NSF. A.S.-N. and Z.W. are supported by the U.S. government through the Office of Naval Research under MURI No. N00014-151-2761, and the National Science Foundation under grant No. ECCS-1708734.
\end{acknowledgments}

\bibliography{heralded_cat_states}
\clearpage


\section*{Supplementary material}

# Painting Non-Classical States of Spin or Motion with Shaped Single Photons: Supplemental Material

*Department of Physics, Stanford University, Stanford, California 94305, USA*

(Dated: June 25, 2018)

In this supplement, we elaborate on details of derivations of the results in the main paper. In Sec. I, we provide analytic expressions for the final heralded state, fidelity $F_{\min}$, and detection rate at arbitrary drive strength in the spin system. In Sec. II we provide detailed derivations of the limits set by the single-atom cooperativity in painting non-classical states of spin or motion in atomic cavity-QED systems.

## I. ANALYTIC RESULTS FOR COLLECTIVE SPIN STATES

The conditional evolution of the spin ensemble or oscillator is governed by the non-Hermitian Hamiltonian

$$H_{\text{eff}} = H_{S/M} + \mathcal{E}c^\dagger + \mathcal{E}^* c + \left(\omega_c - i\frac{\kappa}{2}\right) c^\dagger c, \tag{S1}$$

where $H_{S/M}$ are the Hamiltonians for the spin/oscillator defined in Eqs. 1 and 3 of the main text. From $H_{\text{eff}}$, we can obtain the composite state of the system and cavity field conditioned on no photon having exited the cavity up to time $t$,

$$|\Psi(t)\rangle \equiv e^{-iH_{\text{eff}}t}|\psi_0\rangle \otimes |0\rangle. \tag{S2}$$

Furthermore, conditioned on one photon exiting the cavity at time $t_d$, and no photons exiting before $t_d$, the composite state is given by

$$|\Psi_1\rangle = \sqrt{\kappa}c\hat{\mathcal{T}}e^{-i\int_0^{t_d} H_{\text{eff}}(t)dt}|\psi_0\rangle \otimes |0_c\rangle. \tag{S3}$$

This result is valid for any strength of the drive field.

In the main text, we first proceeded to analyze the conditional dynamics for the limit of a weak drive field, which offers the closest analogy between the spin ensemble and the mechanical oscillator. In the more general case, the spin system is easier to analyze analytically because the atom-light interaction Hamiltonian $\propto c^\dagger c S_z$ commutes with the system Hamiltonian. We here derive the conditional state of the spin system for arbitrary drive strength to arrive at an analytic expression for the fidelity $F_{\min}$.

### A. Heralded Spin State for Arbitrary Drive Strength

In the spin ensemble, we can get the exact analytic form of the time evolution for arbitrary drive strength, because the cavity response depends on the spin state only through $J_z$, which is a constant of motion. To solve for the time evolution, we expand the composite state of the spins and field in the Dicke basis,

$$|\Psi(t)\rangle = \sum_m c_m(t)|m\rangle \otimes |\alpha_m(t)\rangle, \tag{S4}$$

where $|\alpha_m(t)\rangle$ denotes a coherent field of the cavity that depends on time and on $J_z = m$, and $c_m(t)$ is a complex amplitude. The Schrödinger equation with effective Hamiltonian $H_{\text{eff}}$ then yields

$$i\dot{\alpha}_m = \omega_m \alpha_m + \mathcal{E}(t) \tag{S5a}$$

$$\frac{\dot{c}_m}{c_m} = \frac{1}{2}[\dot{\alpha}_m \alpha_m^* + \alpha_m \dot{\alpha}_m^*] - i\mathcal{E}^*(t)\alpha_m, \tag{S5b}$$

where $\omega_m = \omega_c + m\Omega_S - i\frac{\kappa}{2}$ is the complex frequency of the cavity resonance, accounting the finite cavity linewidth and the atom-induced dispersive shift. To arrive at Eqs. S5, we have used the relation

$$\frac{\partial}{\partial t}|\alpha(t)\rangle = -\frac{1}{2}[\dot{\alpha}(t)\alpha^*(t) + \alpha(t)\dot{\alpha}^*(t)]|\alpha(t)\rangle + \dot{\alpha}(t)a^\dagger|\alpha(t)\rangle. \tag{S6}$$

We solve the differential Equations S5 for initial conditions $\alpha_m(0) = 0$ and $c_m(0) = c_m^0$, corresponding to a product state of the atoms in state $|\psi_0\rangle$ and the vacuum field in the cavity. We thus obtain

$$\alpha_m(t) = -i \int_0^t \mathcal{E}(t') e^{-i\omega_m(t-t')} dt', \tag{S7a}$$

$$c_m(t) = c_m^0 \exp\left(\frac{1}{2}|\alpha_m(t)|^2 - i\int_0^t [\mathcal{E}^*(t')\alpha_m] dt'\right). \tag{S7b}$$

For times $t \geq T$, after the drive pulse has ended, we can conveniently express both $\alpha_m(t)$ and the complex amplitude $c_m(t)$ in terms of the Fourier transform of the drive pulse:

$$\tilde{\mathcal{E}}(\omega) = \frac{1}{\sqrt{2\pi}} \int_0^T \mathcal{E}(t) e^{i\omega t} dt. \tag{S8}$$

In terms of $\tilde{\mathcal{E}}$, we have

$$\alpha_m(t) = -i\sqrt{2\pi} e^{-i\omega_m t} \tilde{\mathcal{E}}(\omega_m), \tag{S9a}$$

$$c_m(t) = \exp\left[\frac{1}{2}|\alpha_m(t)|^2 - i\int_{-\infty}^{\infty} \frac{|\tilde{\mathcal{E}}(\omega)|^2}{\omega - \omega_m}\right] c_m^0 \tag{S9b}$$

for $t > T$. Thus, for a given Dicke state $|m\rangle$, the coherent field in the cavity is determined by the component of the drive field at the $m$-dependent complex resonance frequency $\omega_m$. The resulting atom-light entanglement enables the heralded generation of non-classical spin states. In the ideal limit of a weak drive, the complex amplitude $c_m(t)$ is the same as in the initial state, $c_m(t) = c_m^0$, before conditioning on a detected photon. To see the effect of a stronger drive, we now examine the final heralded state for arbitrary drive strength.

Conditioned on detecting a single photon at time $t_d$, the heralded atomic state is

$$|\psi_1(t_d)\rangle = \sqrt{\kappa} \langle 0| \hat{c} |\Psi(t)\rangle = -i\sqrt{2\pi\kappa} \sum_m \xi_m e^{-i\omega_m t_d} \tilde{\mathcal{E}}(\omega_m) c_m^0 |m\rangle, \tag{S10}$$

where

$$\xi_m = \exp\left[-i \int_{-\infty}^{\infty} \frac{|\tilde{\mathcal{E}}(\omega)|^2}{\omega - \omega_m}\right]. \tag{S11}$$

In the limit of a weak drive, $\xi_m = 1$ for all $m$, and Eq. S10 reduces to the recipe of Eq. 14 of the main text for preparing the target state $|\psi_*\rangle$ with coefficients $c_m^f \propto c_m^0 \tilde{\mathcal{E}}(\omega_m)$. At larger drive strength, $\xi_m$ accounts for the effect of light that enters the cavity and leaks back out the input mirror before time $t_d$. This factor can generically reduce the fidelity $F_\epsilon \equiv |\langle\psi_1|\psi_*\rangle|^2 / \langle\psi_1|\psi_1\rangle$ even for perfect detection. However, for the states and drive strengths considered in this paper, $F_\epsilon \approx 1$ to a very good approximation, with the fidelity instead being limited by the effects of imperfect detection considered in Sec. I B.

### B. Fidelity of Heralded Spin States

To determine the fidelity of the final heralded state, we must account for dark counts and finite quantum efficiency in photodetection. Due to these imperfections, we will generically prepare an incoherent mixture of the state produced for a single transmitted photon and states produced when either no photon is transmitted (and we detect a dark count) or multiple photons are transmitted. We calculate the lower bound $F_{\min}$ on the fidelity by assuming that the state conditioned on transmission of a single photon is orthogonal to the states produced when either no photon or more than one photon is transmitted. Then the fidelity is limited by the ratio of successful detector clicks to unsuccessful detector clicks:

$$F_{\min}(t) \equiv \frac{F_\epsilon(t) R_s(t)}{R_t(t) + R_d/Q}, \qquad T < t < t_{\max}, \tag{S12}$$

in terms of the transmission rate $R_t = \kappa \langle c^\dagger c \rangle$, success rate $R_s = \langle \psi_1 | \psi_1 \rangle$, and dark count rate $R_d$.

At the modest drive strengths of interest, where $F_\epsilon \approx 1$ (see Sec. I A), the success rate is approximately

$$R_s(t) \approx \kappa \left|\frac{\epsilon}{\Omega}\right|^2 e^{-|\frac{\epsilon}{\Omega}|^2} e^{-\kappa t}, \qquad T < t < t_{\max}, \tag{S13}$$

where $\epsilon$ parametrizes the drive strength according to Eq. 14 of the main text. The fidelity (Eq. S12) then reduces to

$$F_{\min}(t) \approx \frac{e^{-|\epsilon/\Omega|^2}}{1 + \rho_\epsilon e^{\kappa t}}, \qquad T < t < t_{\max}, \tag{S14}$$

where $\rho_\epsilon = R_d/(Q\kappa |\epsilon/\Omega|^2)$. We use Eq. S14 for plotting $F_{\min}$ in Fig. 2a.

It should be emphasized that $F_{\min}$ is a conservative estimate of the fidelity, and we can also determine the actual fidelity at larger drive strengths. We calculate the atomic state for $n$ transmitted photons, then trace over the detection times of $n-1$ photons. This yields a density matrix for the atoms corresponding to the detection at time $t_d$ of a single heralding photon in the window $T < t < t_{\max}$, with $n-1$ additional transmitted but undetected photons. Roughly, the undetected photons smear the state with extra rotations about the z-axis. The smearing angle $\approx \Omega_S/\kappa$ can be small for $\Omega_S/\kappa \lesssim 1$. In Figure 2c ii, we plotted the Wigner function of the density matrix for a larger drive $\epsilon = \Omega_S$. The fidelity is $F = 0.72$, which is significantly higher than $F_{\min} = 0.37$.

## II. ROLE OF THE COOPERATIVITY IN ATOMIC ENSEMBLES

Finite atom-light coupling strength limits the spin rotation or phase-space displacement that a single photon can induce in an atomic ensemble within the cavity lifetime. In particular, the dispersive atom-light coupling is accompanied by absorption that broadens the cavity linewidth to a value $\kappa_N$ at atom number $N$, decreasing the spin rotation $\Phi_N = \Omega_S/\kappa_N$ or phase-space displacement $D_N \lesssim g_0/\kappa_N$ that the photon imparts within its lifetime $1/\kappa_N$. Fundamental limits on $\Phi_N$ and $D_N$ are set by the single-atom cooperativity $\eta = \mathcal{G}^2/(\kappa\Gamma)$, where $\mathcal{G}$ is the vacuum Rabi frequency and $\Gamma$ is the linewidth of the optical transition to which the cavity mode couples. The average spin rotation produced by a single photon is thus limited to $\Phi_N \leq \sqrt{\eta/(2N)}$, while the average displacement is limited to $D_N \lesssim \sqrt{\eta\omega_r/(2\Omega_M)}$, where $\omega_r$ is the single-atom recoil frequency. Larger rotations or displacements can be attained only with an exponentially decaying success probability. Below, we elaborate on the derivation of these limits.

### A. Atomic Spin Ensembles

To determine the size of the spin rotation that can be induced by a single photon within the cavity lifetime, we consider three-level atoms with ground states $|\downarrow\rangle, |\uparrow\rangle$ and excited state $|e\rangle$. We assume that the cavity mode couples state $|\uparrow\rangle$ to an excited state $|e\rangle$ with vacuum Rabi frequency $g$, and that the other spin state $|\downarrow\rangle$ is unaffected by the light (e.g., the cavity mode is far detuned from transitions involving state $|\downarrow\rangle$). For a detuning $\Delta$ of the resonator mode from the $|\uparrow\rangle \to |e\rangle$ transition, a single photon induces a differential ac Stark shift $\Omega_S = g^2/\Delta$ between the two ground spin states $|\downarrow\rangle$ and $|\uparrow\rangle$.

The interaction time of the photon with the ensemble is limited by the intrinsic loss rate $\kappa_0$ of the cavity and an additional loss rate $(N/2)\Gamma_{\rm sc} = (N/2)(g/\Delta)^2\Gamma$, where $\Gamma$ is the linewidth of the atomic excited state $|e\rangle$, due to the possibility of the photon being scattered out of the resonator by one of the approximately $N/2$ atoms in state $|\uparrow\rangle$. The atomic scattering results in a broadened cavity linewidth

$$\kappa_N = \kappa_0 + \frac{N}{2}\frac{g^2}{\Delta^2}\Gamma = \kappa_0 + \frac{N\Omega^2\Gamma}{2g^2}. \tag{S15}$$

The phase shift $\Omega/\kappa_N$ imparted in the cavity lifetime is maximized by operating at a detuning $\Delta = \Gamma\sqrt{\eta/(2N)}$, where $\eta = 4g^2/(\kappa_0\Gamma)$ is the single-atom cooperativity. Atomic decay then doubles the cavity linewidth ($\kappa_N = 2\kappa_0$), resulting in a ratio $\Omega_S/\kappa_N = \sqrt{\eta/(8N)}$.

The phase shift $\Phi_N \equiv \Omega/\kappa_N$ imparted in the cavity lifetime is an indication of how large a cat can realistically be prepared. Whereas the success probability is approximately constant for $\Phi < \Phi_N$, for larger angles the success probability decays exponentially. Thus, the threshold cooperativity for preparing an $N$-atom cat with phase separation $\Phi$ is $\eta_{\min} \sim N\Phi^2$. More generally, the fidelity of preparing a cat of a given "size" in units of the coherent state width, $\Phi\sqrt{N}$, is set by the single-atom cooperativity and is independent of atom number.

## B. Collective Atomic Motion

The optomechanical interaction Hamiltonian for an ensemble of $N$ atoms linearly coupled to a standing-wave cavity mode is approximately given by [1, 2]

$$H \approx kU_0 c^\dagger c \sum_{i=1}^{N} x_i = NkU_0 c^\dagger c X \equiv g_0 c^\dagger c X/X_0, \tag{S16}$$

where $U_0$ is the AC Stark shift due to a single atom at an antinode of the standing wave, $X$ represents the center-of-mass coordinate, and the zero-point motion $X_0$ of the center of mass is related to oscillator length $x_0$ of a single atom by $X_0 \approx x_0/\sqrt{N}$. Thus, the optomechanical coupling strength is $g_0 = \sqrt{N}U_0\zeta$, where $\zeta = kx_0 = \sqrt{\omega_r/\Omega_M}$ is the Lamb-Dicke parameter set by the single-atom recoil frequency $\omega_r$ at the cavity wavenumber $k$ and by the trap frequency $\Omega_M$.

How large can one make the ratio $g_0/\kappa_N$ that limits the phase-space displacement $D_N$? We first note that the ac Stark shift is $U_0 = \mathcal{G}^2/\Delta$ in terms of the detuning $\Delta$ of the cavity mode from the atomic excited state, i.e., $U_0$ is identical to the value which we called $\Omega_S$ for the spin system in Sec. II A. Thus, the cooperativity sets a limit $\sqrt{N}U_0/\kappa_N \leq \sqrt{\eta/8}$ at an optimal detuning $\Delta/\Gamma = \sqrt{N\eta}/2$ where $\kappa_N = 2\kappa_0$. Hence, $g_0/\kappa_N \leq \zeta\sqrt{\eta/8}$. The Lamb-Dicke parameter $\zeta$ must be small to approximate the sinusoidal potential as a harmonic trap; in seminal experiments by Murch *et al.* [1], $\zeta \approx 0.3$ [1]. Achieving a collective optomechanical coupling $g_0 > \kappa_N$ then requires large *single-atom* cooperativity $\eta \gg 1$. Rewriting the limit on $g_0/\kappa_N$ in terms of the recoil frequency $\omega_r$, we find that the single-photon displacement within the cavity lifetime is limited to $D_N \lesssim \sqrt{\eta\omega_r/(2\Omega_M)}$. Thus, for preparing motional cat states, it is advantageous to operate in a regime of not only high cooperativity but also relatively low trap frequency, which also facilitates meeting the additional requirement $X_1 = g_0/\Omega_M > 1$.

---



\end{document}